\documentclass[12pt]{article}
\usepackage{makeidx,amssymb,amsmath}
\begin{document}

\begin{center}
{\large\bf NOTES ON  DISPERSIONFUL AND DISPERSIONLESS VORTEX FILAMENT EQUATIONS IN
1+1 AND 2+1 DIMENSIONS}

\vspace{2mm}
{\large\it R.Myrzakulov}
\index{Author}
\footnote{
Institute of Physics and Technology, 480082, Alma-Ata, Kazakhstan.
\\E-mail: cnlpmyra@satsun.sci.kz} 

\vspace{2mm}
{\large Institute of Physics and Technology, 480082, Alma-Ata, Kazakhstan
} 
\end{center}

\vspace{2mm}

\abstract{The  vortex filament equations (VFE) in 1+1 and 2+1 dimensions are considered.
Some of these equations are integrable.
Also the VFE with potentials and with self-consistent potentials are presented.
 Finally several examples of integrable dispersionless VFE (dVFE)
are considered.}
\vspace{7mm}
\tableofcontents
\section{Introduction}
The vortex filament equation (VFE) has the form
$${\bf\gamma}_{st}={\bf\gamma}_s \times {\bf\gamma}_{sss},  \eqno(1a)
$$
where $\gamma(s,t)$ denotes the position of the vortex filament in $R^3$ with  $t$
and $s$ being the time and the arclength parameter respectively.
Sometimes we use the following standard form of the VFE
$$
{\bf \gamma}_t={\bf \gamma}_s  \times {\bf \gamma}_{ss} \eqno(1b)
$$
which follows from (1a). Hence we obtain
$$
{\bf \gamma}_{tt}=-\frac{1}{2}({\bf \gamma}_{ss}^{2})_{s} {\bf \gamma}_{s}
-({\bf \gamma}_{ss}^{2}) {\bf \gamma}_{ss}. \eqno(2)
$$

Hasimoto [13] introduced a map $h: \gamma\rightarrow q =ke^{i\int^s \tau(z)dz}, $
in order to transform the VFE into the nonlinear Schrodinger equation (NLSE)
for $q$
$$
iq_t+q_{ss}+2|q|^2q=0. \eqno(3)
$$
Here $k$ and $\tau$ respectively denote the curvature and the torsion along $\gamma$.
 In this paper we consider some dispersionful and dispersionless VFE in 1+1 and 2+1.
Some of these equaions are integrable. Some properties of the VFE from the
the various points
 of view were studied in [1-42].

 \section{Integrable VFE in 1+1}
 First in this section we consider the some well-known (1+1)-dimensional isotropic and anisotropic  VFE. Some examples as follows.

a) The anisotropic VFE. It has the form
 $$
 {\bf\gamma}_{t}={\bf\gamma}_s\times {\bf\gamma}_{ss}+{\bf V},\eqno(4)
    $$
 where ${\bf V}$ is the vector function. As well-known
 this equation is integrable in the
 following cases:

1) For the case ${\bf V}=0$.

2)
$$
 {\bf\gamma}_{t}={\bf\gamma}_s\times {\bf\gamma}_{ss}+{\bf V},
   \eqno(5a)
 $$
  $$
{\bf V}_{s}=\alpha (\gamma^{2}_{ss}){\bf \gamma}_{ss}. \eqno(5b)
 $$

3)

$$
 {\bf\gamma}_{t}={\bf\gamma}_s\times {\bf\gamma}_{ss}+{\bf V},
   \eqno(6a)
 $$
$$
{\bf V}_{s}= {\bf \gamma}_{s}\times A {\bf \gamma}_{s}, \eqno(6b)
$$
 where
 $A=diag(A_1,A_2,A_3), \quad A_{k}=const.
 $

b) The isotropic VFE. It looks like
 $$
 {\bf \gamma}_{st}={\bf \gamma}_s\times {\bf \gamma}_{sss},
   \eqno(7)
 $$
or
 $$
 {\bf \gamma}_t={\bf \gamma}_s\times {\bf \gamma}_{ss}+{\bf f}(t).  \eqno(8)
 $$
Hence as ${\bf f}=0$ we obtain (1b).

c) Next well-known example can be written as [26-27]
$$
 {\bf\gamma}_{st}={\bf\gamma}_{ssss}+\frac{3}{2}({\bf\gamma}_{sss}^2){\bf\gamma}_{ss}.
 \eqno(9)
 $$

d) One of the interesting example is the following VFE [1]
 $$
 {\bf\gamma}_{st}=\alpha({\bf\gamma}_{s}\times {\bf\gamma}_{sss})+
 \beta({\bf\gamma}_{ssss}+\frac{3}{2}({\bf\gamma}_{sss}^2){\bf\gamma}_{ss}).
 \eqno(10)
 $$

e) Finally we present the following known generalization of the VFE
$$
{\bf \gamma}_{st}={\bf \gamma}_s\times {\bf \gamma}_{sss}+\frac{1}{s}{\bf \gamma}_s
\times {\bf \gamma}_{ss}+{\bf \gamma}_s\times A{\bf \gamma}_s.
 \eqno(11)
 $$
 In the isotropic case we have
$$
{\bf \gamma}_{st}={\bf \gamma}_s\times {\bf \gamma}_{sss}+\frac{1}{s}{\bf \gamma}_s
\times {\bf \gamma}_{ss} \eqno(12)
 $$
and so on.

\section{VFE with the potentials}

One of interesting generalizations of the VFE (1) are the VFE with potentials.
May be the simplest example of the such equations is the following anisotropic
Myrzakulov LV (M-LV) equation
 $$
{\bf \gamma}_{st}=((\alpha{\bf \gamma}_{ss}^2+\beta u+\delta)
\gamma_s\times \gamma_{ss})_s+ \gamma_s\times
 A  \gamma_s, \eqno(13)
 $$
 where $u$ is the scalar real  function (potential).
In the isotropic case, the M-LV   equation (13) takes the form
$$
\gamma_t=(\alpha \gamma^2_{ss}+\beta u+\delta)\gamma_{s}\times \gamma_{ss}.  \eqno(14)
$$
 In Table 1 we presented some examples the VFE with potentials. Here and below
$\alpha, \beta, \delta =consts$, $[,]$ is commutator,
$$
g=\mu\gamma_{ss}^2-u+\nu, \quad
\hat\gamma=\gamma\cdot{\bf\sigma},
 \quad {\bf \sigma}=(\sigma_1,\sigma_2,\sigma_3). \eqno(15)
$$
\\
\hfill Table 1.  The VFE with potentials\\
\begin{tabular}{|l|c|}                  \hline
Name of equation & Equation of motion \\ \hline
The M-LVII equation &
$
2i\hat\gamma_{st}=[\hat\gamma_s,\hat\gamma_{sss}]+
u[\hat\gamma_s,\sigma_3] $ \\ \hline
The  M-LVI equation  &
$
2i\hat\gamma_{st}=[\hat\gamma_s,\hat\gamma_{sss}]+
u\gamma_{3s}[\hat\gamma_s,\sigma_3] $ \\ \hline
The M-LV equation  &
$
{\bf \gamma}_{t}=(\mu\gamma_{ss}^2-u+\nu){\bf \gamma}_s\times {\bf \gamma}_{ss}
 $ \\ \hline
The  M-LIV equation &
$
2i\hat\gamma_{st}=n[\hat\gamma_s,\hat\gamma_{sssss}]+2(g
[\hat\gamma_s,
\hat\gamma_{ss}])_s
$ \\ \hline
The M-LIII equation &
$
2i\hat\gamma_{st}=[\hat\gamma_s,\hat\gamma_{sss}]+2iu\hat\gamma_{ss} $ \\ \hline
The M-XCII equation &
$
{\bf \gamma}_{st}=(\alpha {\bf \gamma}_{ss}^{2}+\beta u +\delta)
{\bf \gamma}_{ss} $ \\ \hline
The M-XCIII equation &
$
{\bf \gamma}_{st}=(\alpha \sqrt{{\bf \gamma}_{ss}^{2}}+\beta u +\delta)
{\bf \gamma}_{ss} $ \\ \hline

\end{tabular}
\vspace{0.3cm}

\section{VFE with the self-consistent potentials}

The typical representative of  the VFE with the self-consistent potentials is
the Myrzakulov XLII equation having the form
$$
\gamma_{st}=\{(\mu\gamma^2_{ss} - u +m)\gamma_s \times\gamma_{ss}\}_s+
\gamma_{s}\times A\gamma_{s},\eqno(16a)
$$
$$
u_t+u_s +\lambda ( \gamma^2_{ss})_s = 0. \eqno(16b)
$$
As $A=0$, hence we get the isotropic M-XLII equation
$$
\gamma_{t}=(\mu\gamma^2_{ss} - u +\nu)\gamma_s
 \times\gamma_{ss},\eqno(17a)
$$
$$
u_t+u_s +\lambda ( \gamma^2_{ss})_s = 0. \eqno(17b)
$$

In this section we present some VFE with the self-consistent potentials. Some of
these equations are integrable, e.g. the Myrzakulov XXXIV equation, shortly,
 the M-XXXIV equation (about our notations, see e.g., Refs. [43-52] and also Refs.
 [53-59]).
\\
\hfill
\\
Table 2. \\
\begin{tabular}{|l|c|}                  \hline
Name of equation & Equation of motion\\ \hline
The  M-LII equation &
$
2i\hat\gamma_{st}=[\hat\gamma_s,\hat\gamma_{sss}]+u[\hat\gamma_s,\sigma_3]
$\\
&
$
\rho u_{tt}=\nu^2_0 u_{ss}+\lambda(\hat\gamma_{3s})_{ss}
$ \\  \hline
The M-LI equation &
$
2i\hat\gamma_{st}=[\hat\gamma_s,\hat\gamma_{sss}]+u[\hat\gamma_s,\sigma_3]
$ \\ &
$
\rho u_{tt}=\nu^2_0 u_{ss}+\alpha(u^2)_{ss}+\beta u_{ssss}+
    \lambda(\hat\gamma_{3s})_{ss}
$ \\ \hline
The M-L equation &
$
2i\hat\gamma_{st}=[\hat\gamma_s,\hat\gamma_{sss}]+
u[\hat\gamma_s,\sigma_3]
$ \\ &
$
u_t+u_s+\lambda(\gamma_{3s})_s=0
$ \\ \hline
The M-XLIX equation &
$
2i\hat\gamma_{st}=[\hat\gamma_s,\hat\gamma_{sss}]+u
[\hat\gamma_s,\sigma_3]
$ \\ &
$
u_t+u_s+\alpha(u^2)_s+\beta u_{sss}+\lambda(\hat\gamma_{3s})_s=0
$ \\ \hline
\end{tabular}
\vspace{0.3cm}
\\
\hfill \\Table 3. \\
\begin{tabular}{|l|c|}                  \hline
Name of equation & Equation of motion \\ \hline
The M-XLVIII  equation &
$
2i\hat\gamma_{st}=[\hat\gamma_s,\hat\gamma_{sss}]+
u\gamma_{3s}[\hat\gamma_s,\sigma_3]
$ \\ &
$
\rho u_{tt}=\nu^2_0 u_{ss}+\lambda(\gamma^2_{3s})_{ss}     $ \\ \hline
The M-XLVII equation &
$
2i\hat\gamma_{st}=[\hat\gamma_s,\hat\gamma_{sss}]+
u\gamma_{3s}[\hat\gamma_s,\sigma_3]
$ \\ &
$
\rho u_{tt}=\nu^2_0 u_{ss}+\alpha(u^2)_{ss}+\beta u_{ssss}+
\lambda (\gamma^2_{3s})_{ss}
$ \\ \hline
The M-XLVI equation &
$
2i\hat\gamma_{st}=[\hat\gamma_s,\hat\gamma_{sss}]+
u\gamma_{3s}[\hat\gamma_s,\sigma_3]
$ \\ &
$
u_t+u_s+\lambda(\gamma^2_{3s})_s=0
$ \\ \hline
The M-XLV equation &
$
2i\hat\gamma_{st}=[\hat\gamma_s,\hat\gamma_{sss}]+
u\gamma_{3s}[\hat\gamma_s,\sigma_3]
$ \\ &
$
u_t+u_s+\alpha(u^2)_s+\beta u_{sss}+\lambda(\gamma^2_{3s})_s=0
$\\ \hline
\end{tabular}
\vspace{0.3cm}
\\
\hfill \\Table 4. \\
\begin{tabular}{|l|c|}                  \hline
Name of equation & Equation of motion\\ \hline
The M-XLIV equation &
$
\gamma_{t}=(\mu \gamma^2_{ss} - u +m)\gamma_s\times\gamma_{ss}
$ \\ &
$
\rho u _{tt}=\nu^2_0 u_{ss}+\lambda(\gamma^2_{ss})_{ss}
$ \\ \hline
The M-XLIII equation &
$
\gamma_{t}=(\mu \gamma^2_{ss} - u +m)\gamma_s\times\gamma_{ss}
$ \\ &
$
\rho u _{tt}=\nu^2_0 u_{ss}+\alpha (u^2)_{ss}+\beta u_{ssss}+ \lambda
( \gamma^2_{ss})_{ss}
$ \\ \hline
The M-XLII equation &
$
\gamma_{t}=(\mu \gamma^2_{ss} - u +m)\gamma_s\times\gamma_{ss}
$ \\ &
$
u_t+u_s +\lambda ( \gamma^2_{ss})_s = 0
$ \\ \hline
The M-XLI equation &
$
\gamma_{t}=(\mu \gamma^2_{ss} - u +m)\gamma_s\times\gamma_{ss}
$ \\ &
$
u_t+u_s +\alpha(u^2)_s+\beta u_{sss}+\lambda ( \gamma^2_{ss})_{s} = 0
$ \\ \hline
\end{tabular}
\vspace{0.3cm}
\\
\hfill \\Table 5. \\
\begin{tabular}{|l|c|} \hline
Name of equation & Equation of motion \\ \hline
The  M-XL equation &
$
 2i\hat\gamma_{st}=[\hat\gamma_s,\hat\gamma_{sssss}]+
2\{(\mu \gamma^2_{ss}-u+m)
[\hat\gamma_s,\hat\gamma_{ss}]\}_{s}
$ \\ &
$
\rho u_{tt}=\nu^2_0 u_{ss}+\lambda ( \gamma^2_{ss})_{ss}
$ \\ \hline
The M-XXXIX equation  &
$
2i\hat\gamma_{st}=[\hat\gamma_s,\hat\gamma_{sssss}]+2\{(\mu\gamma^2_{ss}
-u+m)[\hat\gamma_s,\hat\gamma_{ss}]\}_{s}
$ \\ &
$
\rho u_{tt}=\nu^2_0 u_{ss}+\alpha(u^2)_{ss}+\beta u_{ssss}+\lambda
(\gamma^2_{ss})_{ss}
$ \\ \hline
The M-XXXVIII equation &
$
2i\hat\gamma_{st}=[\hat\gamma_s,\hat\gamma_{sssss}]+2\{(\mu\gamma^2_{ss}-u+m)
[\hat\gamma_s,\hat\gamma_{ss}]\}_{s}
$\\ &
$
u_t + u_s + \lambda (\gamma^2_{ss})_s = 0
$ \\ \hline
The M-XXXVII equation &
$
2i\hat\gamma_{st}=[\hat\gamma_s,\hat\gamma_{sssss}]+2\{(\mu \gamma^2_{ss}
-u+m)[\hat\gamma_s,\hat\gamma_{ss}]\}_{s}
$ \\ &
$
u_t + u_s + \alpha(u^2)_s + \beta u_{sss}+\lambda (\gamma^2_{ss})_s = 0
$ \\ \hline
\end{tabular}
\vspace{0.3cm}
\\
\hfill \\Table 6. \\
\begin{tabular}{|l|c|} \hline
Name of equation & Equation of motion\\ \hline
The M-XXXVI equation &
$
2i\hat\gamma_{st}=[\hat\gamma_s,\hat\gamma_{sss}]+2iu\hat\gamma_{ss}
$ \\ &
$
\rho u_{tt}=\nu^2_0 u_{ss}+\frac{\lambda}{4}(tr(\hat\gamma^2_{ss}))_{ss}
$ \\ \hline
The  M-XXXV equation &
$
2i\hat\gamma_{st}=[\hat\gamma_s,\hat\gamma_{sss}]+2iu\hat\gamma_{ss}
$ \\ &
$
\rho u_{tt}=\nu^2_0 u_{ss}+\alpha(u^2)_{ss}+\beta u_{ssss}+
\frac{\lambda}{4}(tr(\hat\gamma^2_{ss}))_{ss}
$ \\ \hline
The M-XXXIV equation &
$
2i\hat\gamma_{st}=[\hat\gamma_s,\hat\gamma_{sss}]+2iu\hat\gamma_{ss}
$\\ &
$
u_t + u_s + \frac{\lambda}{4}(tr(\hat\gamma^2_{ss}))_{s} = 0
$ \\ \hline
The M-XXXIII equation &
$
2i\hat\gamma_{st}=[\hat\gamma_s,\hat\gamma_{sss}]+2iu\hat\gamma_{ss}
$ \\ &
$
u_t + u_s + \alpha(u^2)_s + \beta u_{sss}+\frac{\lambda}{4}(tr(\hat\gamma^2_{ss}))_{s} = 0
$\\ \hline
\end{tabular}
\vspace{0.3cm}
\\
\hfill \\Table 7. \\
\begin{tabular}{|l|c|} \hline
Name of equation & Equation of motion\\ \hline
The M-LXIX equation &
$
{\bf \gamma}_{st}=\frac{1}{\sqrt {{\bf \gamma}^2_{ss}}}(-\sqrt {{\bf \gamma}^2_{ss}
-u^2}{\bf \gamma}_{ss}+u{\bf
\gamma}_s\times{\bf \gamma}_{ss})\hspace{1cm}
$ \\ &
$
u_{s}=v\sqrt {{\bf \gamma}^2_{st}-u^2}
$ \\ &
$ v_t=-{\bf \gamma}_s\cdot ({\bf \gamma}_{st}\times{\bf \gamma}_{ss})
$ \\ \hline
\end{tabular}
\vspace{0.4cm}
\\
\hfill \\ Table 8. \\
\begin{tabular}{|l|c|} \hline
Name of equation & Equation of motion\\ \hline
The M-V equation &
$
\hat\gamma_{t} = \frac{1}{2}[\hat\gamma_s, \hat\gamma_{ss}] + \frac{3}{2}
[\hat\gamma^{2}_s, (\hat\gamma^{2}_{s})_{ss}],\quad
\hat\gamma_s \in osp(2|1)\hspace{0.7cm}
$ \\ \hline
\end{tabular}

\section{VFE with the electromagnetic interaction}
One of interesting problem is the interaction between the vortex
filament and the electromagnetic field. In theory, this interaction
describes by the coupled system of the VFE and the Maxwell equations.
In the soliton limit, hence, we get the coupled system of the VFE
and the Schrodinger-type equation. As example, we consider the
following system of the  coupled equations
$$
\gamma_{st}=[(\alpha |\phi|^2+\beta \gamma^2_{ss}+\delta)\gamma_s
\times \gamma_{ss}]_s+\gamma_s\times A \gamma_{s}, \eqno(18a)
$$
$$
i\phi_t+\phi_{ss}+(\mu|\phi|^2+\nu\gamma^2_{ss}+\lambda)\phi=0. \eqno(18b)
$$
Hence in the isotropic case we have
$$
\gamma_{t}=(\alpha |\phi|^2+\beta \gamma^2_{ss}+\delta)\gamma_s
\times \gamma_{ss}, \eqno(19a)
$$
$$
i\phi_t+\phi_{ss}+(\mu|\phi|^2+\nu\gamma^2_{ss}+\lambda)\phi=0. \eqno(19b)
$$
In this section we present some systems of equations which describe interaction between the vortex filament
and electromagnetic fields.
\\
\hfill
\\Table 9. \\
\begin{tabular}{|l|c|} \hline
Name of equation & Equation of motion \\ \hline
The  M-LXXI equation &
$
 {\bf \gamma}_{st}={\bf\gamma}_s\times{\bf\gamma}_{sss}+
\alpha|\phi|^2{\bf\gamma}_{ss}+{\bf\gamma}_s
\times A{\bf\gamma}_s
$ \\ &
$
i\phi_t+\phi_{ss}+\lambda {\bf \gamma}^2_{ss}\phi=0
$ \\ \hline
The M-LXXII equation  &
$
{\bf\gamma}_{st}={\bf\gamma}_s\times{\bf\gamma}_{sss}+\alpha|\phi|^2{\bf\gamma}_{ss}
+{\bf\gamma}_s
\times A{\bf\gamma}_s
$ \\ &
$
i\phi_t+\phi_{ss}+i\lambda ({\bf \gamma}^2_{ss}\phi)_s=0
$ \\ \hline
The M-LXXIII equation &
$
{\bf\gamma}_{st}={\bf\gamma}_s\times{\bf\gamma}_{sss}+\alpha|\phi|^2{\bf\gamma}_{ss}
+{\bf\gamma}_s
\times A{\bf\gamma}_s
$\\ &
$
i\phi_t+\phi_{ss}+i\lambda {\bf \gamma}^2_{ss}\phi_s=0
$ \\ \hline
\end{tabular}
\vspace{0.3cm}
\\
\hfill \\Table 10. \\
\begin{tabular}{|l|c|} \hline
Name of equation & Equation of motion \\ \hline
The  M-LXXIV equation &
$
 {\bf \gamma}_{t}=(\mu|\phi|^2+\nu){\bf\gamma}_s\times{\bf\gamma}_{ss}
$ \\ &
$
i\phi_t+\phi_{ss}+\lambda {\bf \gamma}^2_{ss}\phi=0
$ \\ \hline
The M-LXXV equation  &
$
{\bf \gamma}_{t}=(\mu|\phi|^2+\nu){\bf\gamma}_s\times{\bf\gamma}_{ss}
$ \\ &
$
i\phi_t+\phi_{ss}+i\lambda ({\bf \gamma}^2_{ss}\phi)_s=0
$ \\ \hline
The M-LXXVI equation &
$
{\bf \gamma}_{t}=(\mu|\phi|^2+\nu){\bf\gamma}_s\times{\bf\gamma}_{ss}
$\\ &
$
i\phi_t+\phi_{ss}+i\lambda {\bf \gamma}^2_{ss}\phi_s=0
$ \\ \hline
\end{tabular}
\vspace{0.3cm}
\\
\hfill \\Table 11. \\
\begin{tabular}{|l|c|} \hline
Name of equation & Equation of motion \\ \hline
The  M-LXXVII equation &
$
{\bf \gamma}_{st}=\alpha{\bf\gamma}_s\times {\bf \gamma}_{sssss}+
\{(\mu|\phi|^2+\nu){\bf\gamma}_s\times{\bf\gamma}_{ss}\}_s
$ \\ &
$
i\phi_t+\phi_{ss}+\lambda {\bf \gamma}^2_{ss}\phi=0
$ \\ \hline
The M-LXXVIII equation  &
$
{\bf \gamma}_{st}=\alpha{\bf\gamma}_s\times {\bf \gamma}_{sssss}+
\{(\mu|\phi|^2+\nu){\bf\gamma}_s\times{\bf\gamma}_{ss}\}_s
$ \\ &
$
i\phi_t+\phi_{ss}+i\lambda ({\bf \gamma}^2_{ss}\phi)_s=0
$ \\ \hline
The M-LXXIX equation &
$
{\bf \gamma}_{st}=\alpha{\bf\gamma}_s\times {\bf \gamma}_{sssss}+
\{(\mu|\phi|^2+\nu){\bf\gamma}_s\times{\bf\gamma}_{ss}\}_s
$\\ &
$
i\phi_t+\phi_{ss}+i\lambda {\bf \gamma}^2_{ss}\phi_s=0
$ \\ \hline
\end{tabular}
\vspace{0.3cm}

\section{Integrable VFE in 2+1}
It is well-known that each (1+1)-dimensional integrable systems admits several (not one)
integrable (and not integrable) systems in 2+1 dimensions.
In the previous sections we presented some examples integrable and nonintegrable VFE in 1+1 dimensions. In this
section we consider the several VFE in 2+1 dimensions which are the (2+1)-dimensional integrable extensions of the
VFE (1) or (4). Some examples as follows.

i) The anisotropic (2+1)-dimensional VFE.
$$
{\bf \gamma}_{st}={\bf \gamma}_s\times({\bf \gamma}_{sss}+\alpha^2
{\bf \gamma}_{syy})+u_s {\bf \gamma}_{sy}+u_y {\bf \gamma}_{ss}+W_s,
\eqno(20a)
$$
$$
u_{ss}-\alpha^2u_{yy}=-2\alpha^2 {\bf\gamma}_s({\bf\gamma}_{ss}\times {\bf\gamma}_{sy}),
\eqno(20b)
$$
$$
W_y=F_s.  \eqno(20c)
$$
Hence we obtain the well-known isotropic version which has the form
$$
{\bf \gamma}_{st}={\bf \gamma}_s\times ({\bf \gamma}_{sss}+\alpha^2 {\bf \gamma}_{syy})+
u_s {\bf \gamma}_{sy}+u_y {\bf \gamma}_{ss},
\eqno(21a)
$$
$$
u_{ss}-\alpha^2u_{yy}=-2\alpha^2 {\bf\gamma}_s\cdot({\bf\gamma}_{ss}\times {\bf\gamma}_{sy}).
 \eqno(21b)
$$

ii) The anisotropic Myrzakulov I equation (about our  notations, see e.g., Refs [43-52] and also [53-59]).
It reads as [44]

$$
{\bf \gamma}_{st}=({\bf \gamma}_s\times {\bf \gamma}_{sy}+u{\bf \gamma}_s)_s
+ \gamma_s \times V,  \eqno(22a)
$$
$$
u_s=-{\bf \gamma}_s\cdot({\bf \gamma}_{ss}\times {\bf \gamma}_{sy}),  \eqno(22b)
$$
$$
V_y=A \gamma_{sy}.  \eqno(22c)
$$
In the isotropic case we get (the isotropic M-I equation)
$$
{\bf \gamma}_{st}=({\bf \gamma}_s\times {\bf \gamma}_{sy}+u{\bf \gamma}_s)_s,  \eqno(23a)
$$
$$
u_s=-{\bf \gamma}_s\cdot({\bf \gamma}_{ss}\times {\bf \gamma}_{sy}),  \eqno(23b)
$$
or
$$
{\bf \gamma}_{t}={\bf \gamma}_{s} \times {\bf \gamma}_{sy} + u{\bf \gamma}_{s},  \eqno(24a)
$$
$$
u_s=-{\bf \gamma}_s\cdot({\bf \gamma}_{ss} \times \gamma_{sy}).  \eqno(24b)
$$

iii) The Myrzakulov II equation [44]
$$
{\bf\gamma}_{st}=({\bf\gamma}_s\times{\bf\gamma}_{sy}+u{\bf\gamma}_s)_s+2cb^2{\bf\gamma}_{sy}-
4c\upsilon{\bf\gamma}_{ss},  \eqno(25a)
$$
$$
u_s=-{\bf\gamma}_s\cdot({\bf\gamma}_{ss}\times{\bf\gamma}_{sy}),  \eqno(25b)
$$
$$
\upsilon_s=\frac{1}{16b^2c^2}({\bf\gamma}^2_{ss})_y.  \eqno(25c)
$$

iv) The Myrzakulov III equation [44]
$$
{\bf\gamma}_{st}=({\bf\gamma}_s\times{\bf\gamma}_{sy}+u{\bf\gamma}_s)_s+
2b(cb+d){\bf\gamma}_{sy}-
4c\upsilon{\bf\gamma}_{ss},  \eqno(26a)
$$
$$
u_s=-{\bf\gamma}_s\cdot({\bf\gamma}_{ss}\times{\bf\gamma}_{sy}),  \eqno(26b)
$$
$$
\upsilon_s=\frac{1}{4(2bc+d)^2}({\bf\gamma}^2_{ss})_y.  \eqno(26c)
$$

v) The Myrzakulov XXII equation [44]
$$
-i\gamma_{st}=\frac{1}{2}([\gamma_s,\gamma_{sy}]+2iu\gamma_s)_{s}+\frac{i}{2}v
\gamma_{ss}-2ia^2\gamma_{sy},  \eqno(27a)
$$
$$
u_s=-{\bf\gamma_s}\cdot({\bf\gamma}_{ss}\times {\bf\gamma}_{sy}), \eqno(27b)
$$
$$
v_s=\frac{1}{4a^2}({\bf\gamma}^2_{ss})_y.  \eqno(27c)
$$

vi) The Myrzakulov VIII equation [44]

$$
{\bf \gamma}_{st}={\bf \gamma}_s\times {\bf \gamma}_{sss}
+u{\bf \gamma}_{ss}+W_s,  \eqno(28a)
$$
$$
u_y={\bf\gamma}_s\cdot({\bf\gamma}_{ss}\times {\bf\gamma}_{sy}),  \eqno(28b)
$$
$$
W_y=F_x.  \eqno(28c)
$$
In the isotropic case, we obtain
$$
{\bf \gamma}_{st}={\bf \gamma}_s\times {\bf \gamma}_{sss}+u {\bf \gamma}_{ss},  \eqno(29a)
$$
$$
 u_y={\bf\gamma}_s\cdot({\bf\gamma}_{ss}\times {\bf\gamma}_{sy}). \eqno(29b)
$$

vii) The Myrzakulov XX equation [44]
$$
{\bf\gamma}_{st}+{\bf\gamma}_s\times\{(b+1){\bf\gamma}_{sss}-b{\bf\gamma}_{syy}+
bu_y{\bf\gamma}_{sy}+(b+1)u_s{\bf\gamma}_{ss}=0,  \eqno(30a)
$$
$$
u_{sy}={\bf\gamma}_s\cdot({\bf\gamma}_{ss}\times{\bf\gamma}_{sy}).  \eqno(30b)
$$

viii) The Myrzakulov IX equation [44]
$$
i\hat\gamma_{st}+\frac{1}{2}[\hat\gamma_s,M_1\hat\gamma_s]
+A_2\hat\gamma_{ss}+A_1\hat\gamma_{sy}=0,
\eqno(31a)
$$
$$
M_2u=\frac{\alpha^2}{2i}tr(\hat\gamma_s[\hat\gamma_{ss},
\hat\gamma_{sy}]). \eqno(31b)
$$

Finally we note that all of these (2+1)-dimensional VFE are integrable.
And in 1+1 dimensions
they reduce to the VFE (1) or (4). Of course, there are exist also some
nonintegrable (2+1)-dimensional extensions of the VFE (1) or (4).
One of such
extensions has the form
$$
{\bf\gamma}_{st}={\bf\gamma}_s\times({\bf\gamma}_{sss}+b{\bf\gamma}_{syy})+
{\bf\gamma}_{s}\times A{\bf\gamma}_{s}.  \eqno(32)
$$
The isotropic version of the (32) has the form
$$
{\bf\gamma}_{st}={\bf\gamma}_s\times({\bf\gamma}_{sss}
+b{\bf\gamma}_{syy}).  \eqno(33)
$$

\section{Integrable  planar VFE}

In this section we present some planar filament equations. Here $\gamma(s,t)$ denotes
an evolving planar curve, parametrized by arclength $s, k $ is its curvature. Such
equations have been studied from the different point of views (see, for example,
Ref.  [27]).

Example 1. First we consider the following planar VFE [26-27]
$$
{\bf \gamma}_{st}={\bf \gamma}_{ssss}+a{\bf \gamma}_{ss}+b{\bf \gamma}_s,
 \eqno(34)
$$
where
$$
 a= \gamma_{ss}^2+
\frac{3}{4}\sqrt{\gamma_{ss}^2}, \quad b=\frac{3}{2}(\gamma^2_{ss})_s.
\eqno(35)
$$

Example 2.  The Myrzakulov X equation.
It is integrable and has the form [44]
$$
\gamma_{st}+ \gamma_{ssss}+3\sqrt{\gamma^2_{ss}}\gamma_{ss}-
3\alpha^2 \gamma_{yy}=0.      \eqno(36)
$$

Finally we note that the equations (34) and (36) are  integrable.

\section{Integrable dispersionless VFE}

A considerable interest has been paid recently to dispersionless or quasi-classical
limits of integrable equations and hierarchies. Study of dispersionless hierarchies is of great
importance since they arise in the analysis of various problems in physics, mathematics
and applied mathematics from the theory of quantum fields and strings to the theory
of conformal maps on the complex plane.

Above we presented some despersionful VFE.
Now we want present some examples integrable dispersionless VFE (dVFE) in 1+1 and 2+1
dimensions. For simplicity, we consider only the planar dVFE.

Example 1. Simplest example integrable dVFE reads as
$$
\gamma_{st}=\frac{3}{4}\sqrt{\gamma^2_{ss}}\gamma_{ss}. \eqno(37)
$$
It is the Myrzakulov XCVIII equation [44]. As well-known it is L-equivalent to the
dispersionless KdV (dKdV)
equation (or the Riemann equation)
$$
k_t=\frac{3}{2}kk_s,  \eqno(38)
$$
where $k$ is the curvature of the plane curve.

Example 2. The Myrzakulov XCVII equation. It is integrable and has the form [44]
$$
(\gamma_{st}-\frac{3}{4}\sqrt{\gamma^2_{ss}}\gamma_{ss})_s=-\frac{3}{4i}
(\gamma_{syy}\cdot
\sigma_2\gamma_s),  \eqno(39)
$$
where
$$
\sigma_2=
\left( \begin{array}{cc}
0 & -i \\
i & 0
\end{array} \right). \eqno(40)
$$

Example 3. The Myrzakulov XCVI equation. This equation is also integrable. It looks like [44]
$$
{\bf \gamma}_{st}=[W-3\partial^{-1}_{\bar z}(\sqrt{{\bf \gamma}^2_{sz}})_z]{\bf\gamma}_{sz},
\eqno(41a)
$$
$$
W_z=-3[\sqrt{{\bf\gamma}^2_{sz}}\partial^{-1}_z(\sqrt{{\bf\gamma}^2_{sz}})_{\bar z}]_{\bar z},
\eqno(41b)
$$
where $  z=s+iy $.

Example 4.  The Myrzakulov XCV equation,
which is integrable and reads as [44]
$$
\gamma_{st}=(\frac{3}{4}V-\frac{1}{2}\gamma^2_{ss}+W)\gamma_{ss}, \eqno(42a)
$$
$$
V_s=(\sqrt{\gamma^2_{ss}})_y, \eqno(42b)
$$
$$
(W\sqrt{\gamma^2_{ss}})_s=(\frac{3}{4}V_y-\frac{3}{2}\gamma^2_{ss})_y.  \eqno(42c)
$$

Example 5.  The Myrzakulov C equation which reads as [44, 60]
$$
\gamma_{st}=f_{1}{\bf \gamma}_{s}\times {\bf \gamma}_{ss}+f_{2}{\bf \gamma}_{ss}
+f_3 \gamma\times \gamma_s, \eqno(43)
$$
where $f_{k}({\bf \gamma}, {\bf \gamma}_{s}, ... )$ is some scalar
 functions of the arguments. Note that   the M-C equation (43)
 is L-equivalent to the Benney equation.
Note that these dVFE are related with the integrable dispersionless
spin systems (see, e.g. Ref. [43]).

\section{Conclusion}

In this paper we have presented some dispersionful and dispersionless VFE in 1+1 and
2+1 dimensions. Some of these equations are integrable. All of these equations admit
different types exact solutions like solitons, knotes, breaking waves, etc. It is of
great interest to study such solutions of the VFE in 1+1 and 2+1 dimensions and their
integrability. We are currently investigating this issue and our finding
will appear in a future paper.

\end{document}